\documentclass[aps,pra,preprint,showpacs]{revtex4}
\usepackage{pstricks}
\usepackage{graphicx}
\begin{document}
\preprint{Version 4.0}

\title{Two-center two-electron integrals with exponential functions}

\author{Krzysztof Pachucki}
\email[]{krp@fuw.edu.pl}

\affiliation{Institute of Theoretical Physics, University of Warsaw,
             Ho\.{z}a 69, 00-681 Warsaw, Poland}

\begin{abstract}
We present an efficient approach to evaluate 
two-center two-electron integrals with exponential functions
and with an arbitrary polynomial in electron-nucleus 
and electron-electron distances. We show that 
the master integral with the single negative power of all distances
can be obtained from the second order differential 
equation in $r$, the distance between nuclei.
For particular values of nonlinear parameters corresponding to
the James-Coolidge basis, we find a fully analytic expression.
For integrals with arbitrary powers of all distances,
we construct recursion relations which starts
from the master integral. The presented approach opens a window
for the high precision calculations of relativistic effects 
in diatomic molecules.
\end{abstract}

\maketitle
\section{Introduction}
In order to calculate accurately relativistic effects in atomic or
molecular systems, the wave function should satisfy cusp conditions.
Namely, at the electron-nucleus and electron-electron coalescence points
the derivative of the wave function is proportional to the wave function itself,
with coefficients proportional to $-Z$ or $1/2$ correspondingly.
The widely used Gaussian basis set does not satisfy any of these conditions,
therefore the numerical convergence of relativistic effects, is slow
or sometimes even does not lead to the right result. It has been found
recently \cite{f12} that inclusion of the single factor $e^{-\lambda\,r_{12}}$ 
on the top of Gaussian functions significantly improves convergence of the
nonrelativistic energy. Even better analytic properties are satisfied 
by the exponential (Slater) basis set
with polynomials of electron-nucleus and electron-electron distances.
Moreover, the large distance asymptotics of exponential functions
agrees with that obtained from quasi-classical expansion. However
the usage of exponential functions in molecular
calculations  has been limited due to inherent difficulties in the
accurate and efficient calculations of multi-center integrals. In this work we
overcome this problem for the simplest case of the two-electron and
two-center integral, with arbitrary nonlinear parameters and
arbitrary powers of electron-nucleus and electron-electron
distances. We show, that the master integral, with single negative power
of all distances satisfies a second order differential equation
in the nuclear distance $r$. This equation can be solved numerically,
or in the case of James-Coolidge basis, is solved analytically
in terms of Ei- the exponential integral functions. The integrals
with higher powers of electron distances are obtained by analytic
recursion relations which involve the master integral and
elementary functions.  Our approach is based on a set of
integration by parts identities, and is well established in the area
of multi-loop Feynman diagrams \cite{fdiag}. 
Integration by parts identities, similar to those derived here,
were recently applied to three-electron one-center Hylleraas
\cite{lit} and exponentially correlated \cite{harris} integrals.

\begin{figure}[!htb]
\begin{pspicture}(-5,-3)(5,5)

\psdot*[dotscale=1](0,0)
\psdot*[dotscale=1](0,4)
\psdot*[dotscale=1](3.4641,-2)
\psdot*[dotscale=1](-3.4641,-2)

\psline[linewidth=1pt](0,4)
\psline[linewidth=1pt](3.4641,-2)
\psline[linewidth=1pt](-3.4641,-2)

\psline[linewidth=1pt](0,4)(3.4641,-2)
\psline[linewidth=1pt](3.4641,-2)(-3.4641,-2)
\psline[linewidth=1pt](-3.4641,-2)(0,4)

\rput(-0.2,0.2){$0$}
\rput(0.3,0.2){e$_2$}
\rput(0.1,4.4){$1\,$e$_1$}
\rput(-3.6,-2.3){$2$ $A$}
\rput(3.6,-2.3){$3$ $B$}

\rput(0.3,1.8){$w_1$}
\rput(-1.5,-1.2){$w_2$}
\rput(1.5,-1.2){$w_3$}

\rput(0.0,-2.3){$u_1$}
\rput(-2.1,1.0){$u_3$}
\rput(2.1, 1.0){$u_2$}

\end{pspicture}
\caption{\label{figure1} The tetrahedron geometry of the master
integral. The double notation is used for vertices, namely
vertices $1\, ({\rm e}_1)$ and $0\, ({\rm e}_2)$ correspond to the position of the
first and the second electron, vertices $2\, (A)$ and $3\, (B)$ correspond to
the position of nuclei. The nonlinear parameter $w_1$ is related to
$r_{12}$ distance and $u_1=t$ to $r=r_{AB}$.}
\end{figure}
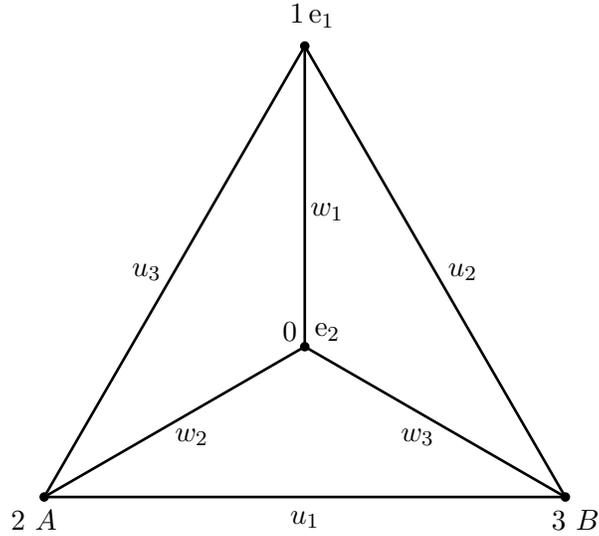

Let us now define the master two-electron and two-center integral $f(r)$,
where $r=r_{AB}$,
\begin{equation}
f(r) = \int \frac{d^3 r_1}{4\,\pi}\,\int \frac{d^3 r_2}{4\,\pi}\,
\frac{e^{-u_3\,r_{1A}}}{r_{1A}}\,
\frac{e^{-u_2\,r_{1B}}}{r_{1B}}\,
\frac{e^{-w_2\,r_{2A}}}{r_{2A}}\,
\frac{e^{-w_3\,r_{2B}}}{r_{2B}}\,\frac{r}{r_{12}},\label{01}
\end{equation}
where $1,2$ are the positions of the electrons, $A,B$ positions of the nuclei,
and the notation for nonlinear parameters will be clarified later
on. For practical reasons, it is easier at first to consider another
integral $g$ defined by
\begin{equation}
g(u_1) = \int \frac{d^3 \rho_1}{4\,\pi}\,\int \frac{d^3 \rho_2}{4\,\pi}\,\int
\frac{d^3 \rho_3}{4\,\pi} \,\frac{e^{-w_1\,\rho_1-w_2\,\rho_2-w_3\,\rho_3
 -u_1\,\rho_{23}-u_2\,\rho_{31}-u_3\,\rho_{12}}}{
\rho_{23}\,\rho_{31}\,\rho_{12}\,\rho_{1}\,\rho_{2}\,\rho_{3}},\label{02}
\end{equation}
with $\vec\rho_1=\vec r_{12},\,\vec\rho_2=\vec r_{2A},\,\vec\rho_3=\vec r_{2B}$,
which is related to $f$ by a Laplace transform, namely
\begin{eqnarray}
g(t)\Bigr|_{w_1=0} &=&
\int\frac{d^3r}{4\,\pi}\,f(r)\,\frac{e^{-t\,r}}{r^2} = \int_0^\infty dr\,f(r)\,e^{-t\,r},\label{03}
\\
f(r) &=&
\frac{1}{2\,\pi\,i}\,\int_{-i\,\infty+\epsilon}^{i\,\infty+\epsilon}dt\,e^{t\,r}\,g(t)\Bigr|_{w_1=0}.
\label{04}
\end{eqnarray}
The integral $f_n$ ($f_0=f$) with the integer power $n\geq 0$ of inter-electronic distance $r_{12}$
\begin{equation}
f_n(r) = \int \frac{d^3 r_1}{4\,\pi}\,\int \frac{d^3 r_2}{4\,\pi}\,
\frac{e^{-u_3\,r_{1A}}}{r_{1A}}\,
\frac{e^{-u_2\,r_{1B}}}{r_{1B}}\,
\frac{e^{-w_2\,r_{2A}}}{r_{2A}}\,
\frac{e^{-w_3\,r_{2B}}}{r_{2B}}\,r_{12}^{n-1}\,r,\label{05}
\end{equation}
can be obtained from $g$ by differentiation over $w_1$, see Eq.
(\ref{40}). A similar situation holds for integrals with integer powers $i,j,k,l
\geq 0$ of all electron-nucleus distances
\begin{equation}
f_n(i,j,k,l;r) = \int \frac{d^3 r_1}{4\,\pi}\,\int \frac{d^3 r_2}{4\,\pi}\,
\frac{e^{-u_3\,r_{1A}}}{r_{1A}^{1-i}}\,
\frac{e^{-u_2\,r_{1B}}}{r_{1B}^{1-j}}\,
\frac{e^{-w_2\,r_{2A}}}{r_{2A}^{1-k}}\,
\frac{e^{-w_3\,r_{2B}}}{r_{2B}^{1-l}}\,\frac{r}{r_{12}^{1-n}}.\label{06}
\end{equation}
They can be obtained by further differentiation of $g$ in Eq. (\ref{02})
over nonlinear parameters $u_2, u_3, w_2,w_3$. For this we derive recursion 
relations, which make possible in practice the evaluation
of integrals with many powers of electron-electron and electron-nucleus
distances. The function $g$ will be
calculated from the pertinent integral in the momentum space, namely
\begin{equation}
g(u_1) = G(1,1,1;1,1,1),\label{07}
\end{equation}
where
\begin{eqnarray}
G(m_1,m_2,m_3;m_4,m_5,m_6) &=& \frac{1}{8\,\pi^6}\,\int d^3k_1\int d^3k_2\int d^3k_3\,
(k_1^2+u_1^2)^{-m_1}\,(k_2^2+u_2^2)^{-m_2} \nonumber \\ &&(k_3^2+u_3^2)^{-m_3}\,
(k_{32}^2+w_1^2)^{-m_4}\,(k_{13}^2+w_2^2)^{-m_5}\,(k_{21}^2+w_3^2)^{-m_6}.\label{08}
\end{eqnarray}
The topology of these integrals and the notation is presented in Fig. (\ref{figure1}).

In Sec. II we will derive a differential equation which is satisfied
by the function $g$. This equation can be used to obtain an analytic form of $g$.
However, we find that it is too complicated for practical applications. 
In Sec. III we perform an inverse Laplace transform to obtain
a differential equation for the master integral $f(r)$, which can be
conveniently solved numerically. In Sec. IV and V, using 
this differential equation for $g$, we derive recursion relations for
evaluation of $f_n(i,j,k,l;r)$ in Eq. (\ref{06}). In Sec. VI and VII 
we work out special cases of direct and exchange atomic integrals.
In Sec. VIII we consider symmetric integrals
which are suited for the H$_2$ molecule, and for them we obtain 
a simple analytic form. Finally in Sec. IX we present a short summary.

\section{Differential equation}
We use the method of integration by parts
identities~\cite{fdiag}, which is by now standard in the
analytical calculation of Feynman diagrams. In our case, it amounts
to considering the following 9 identities in the momentum space
representation of the integral $G$, $(i,j=1,2,3)$
\begin{eqnarray}
&&0 \equiv {\rm id}(i,j) =
\int d^3k_1\int d^3k_2\int d^3k_3\,\frac{\partial}{\partial\,{\vec k_j}}
 \Bigl[ \vec k_i\,(k_1^2+u_1^2)^{-m_1}
\nonumber \\ &&
(k_2^2+u_2^2)^{-m_2}\,(k_3^2+u_3^2)^{-m_3}
(k_{32}^2+w_1^2)^{-m_4}\,(k_{13}^2+w_2^2)^{-m_5}\,(k_{21}^2+w_3^2)^{-m_6}
\Bigr] ,
\label{09}
\end{eqnarray}
which are trivially valid, because the integral of the derivative of a function
vanishing at infinity vanishes. These identities have been first introduced in
Ref. \cite{lit} for the calculation of one-center three-electron Hylleraas integrals.
They group naturally into three sets. The first set consists of
id$(1,1)$, id$(2,1)$, and id$(3,1)$. Other sets are obtained by changing
the second argument from $1$ into $2$ or $3$.
The reduction of the scalar products from the numerator leads to 
identities between functions $G$ of different arguments. Whenever
$m_i=0$,  $G$ becomes a known three-body integral, for example
\begin{eqnarray}
G(0,1,1;1,1,1) &=& \frac{1}{2\,w_1}\,\biggl[
  {\rm Li}\biggl(1 - \frac{u_2 + u_3 + w_2 + w_3}{u_2 + u_3 + w_1}\biggr)
+ {\rm Li}\biggl(1 - \frac{u_2 + u_3 + w_2 + w_3}{w_1 + w_2 + w_3}\biggr)\nonumber \\ &&
 + \frac{1}{2}\,\ln^2\biggl(\frac{w_1 + w_2 + w_3}{u_2 + u_3 + w_1}\biggr) + \frac{\pi^2}{6}\biggr],
\label{10}
\end{eqnarray}
where Li is the dilogarithmic function. If we assume all $m_i=1$ in
Eq. (\ref{08}) and solve an arbitrary set of three equations, for
example the last one, against three unknowns $G(1,1,2;1,1,1),
G(1,1,1;2,1,1), G(1,1,1;1,2,1)$, then the solution for
$G(1,1,1;2,1,1)$ is of the form
\begin{equation}
\frac{1}{2}\,\frac{\partial\sigma}{\partial w_1}\,G(1,1,1;1,1,1)
-2\,w_1\,\sigma\,G(1,1,1;2,1,1) + P(w_1,u_1;w_2,u_2;w_3,u_3) = 0,\label{11}
\end{equation}
where $\sigma$ is a polynomial with a tetrahedral symmetry
\begin{eqnarray}
\sigma &=&
u_1^2\,u_2^2\,w_3^2  + u_2^2\,u_3^2\,w_1^2  + u_1^2\,u_3^2\,w_2^2 + w_1^2\,w_2^2\,w_3^2
 + u_1^2\,w_1^2\,(u_1^2+w_1^2-u_2^2-u_3^2-w_2^2-w_3^2)
\nonumber \\ &&
 + u_2^2\,w_2^2\,(u_2^2+w_2^2-u_1^2-u_3^2-w_1^2-w_3^2)
 + u_3^2\,w_3^2\,(u_3^2+w_3^2-u_2^2-u_1^2-w_1^2-w_2^2),\label{12}
\end{eqnarray}
and
\begin{eqnarray}
    & &  P(w_1,u_1;w_2,u_2;w_3,u_3) \nonumber \\
&=& u_1\,w_1\,[(u_1 + w_2)^2 - u_3^2]\,\Gamma(u_2+w_1,u_3,u_1+w_2)
\nonumber \\ &&
   +u_1\,w_1\,[(u_1 + u_3)^2 - w_2^2]\,\Gamma(w_1+w_3,w_2,u_1+u_3)
\nonumber \\ &&
   -[u_1^2\,w_1^2 + u_2^2\,w_2^2 - u_3^2\,w_3^2 + w_1\,w_2\,(u_1^2 + u_2^2 - w_3^2)]
    \,\Gamma(u_1+u_2,w_3,w_1+w_2)
\nonumber \\ &&
   -[u_1^2\,w_1^2 - u_2^2\,w_2^2 + u_3^2\,w_3^2 + w_1\,w_3\,(u_1^2 + u_3^2 - w_2^2)]
    \,\Gamma(u_1+u_3,w_2,w_1+w_3)
\nonumber \\ &&
   +[u_2\,(u_2 + w_1)\,(u_1^2 + u_3^2 - w_2^2) - u_3^2\,(u_1^2 + u_2^2 - w_3^2)]
    \,\Gamma(u_1+w_2,u_3,u_2+w_1)
\nonumber \\ &&
   +[u_3\,(u_3 + w_1)\,(u_1^2 + u_2^2 - w_3^2) - u_2^2\,(u_1^2 + u_3^2 - w_2^2)]
    \,\Gamma(u_1+w_3,u_2,u_3+w_1)
\nonumber \\ &&
   -w_1\,[w_2\,(u_1^2 - u_2^2 + w_3^2) + w_3\,(u_1^2 + w_2^2 - u_3^2) ]
    \,\Gamma(u_2+u_3,w_1,w_2+w_3)
\nonumber \\ &&
   -w_1\,[u_2\,(u_1^2 - w_2^2 + u_3^2) + u_3\,(u_1^2 + u_2^2 - w_3^2)]
    \,\Gamma(w_2+w_3,w_1,u_2+u_3),\label{13}
\end{eqnarray}
with
\begin{equation}
\Gamma(\alpha_1,\alpha_2,\alpha_3)=\frac{\ln(\alpha_1+\alpha_2)-\ln(\alpha_1+\alpha_3)}
{(\alpha_2-\alpha_3)\,(\alpha_2+\alpha_3)}.
\end{equation}
Since $G(1,1,1;1,1,1) = g$ and
\begin{equation}
 G(1,1,1;2,1,1) =
-\frac{1}{2\,w_1}\,\frac{\partial g}{\partial w_1},\label{14}
\end{equation}
Eq. (\ref{11}) takes the form of a differential equation
\begin{equation}
\sigma\,\frac{\partial g}{\partial w_1} +
\frac{1}{2}\,\frac{\partial\sigma}{\partial w_1}\,g +
P(w_1,u_1;w_2,u_2;w_3,u_3) = 0\,,\label{15}
\end{equation}
or
\begin{equation}
\sqrt{\sigma}\frac{\partial}{\partial w_1}(\sqrt{\sigma}\,g) +
P(w_1,u_1;w_2,u_2;w_3,u_3) = 0\,.\label{16}
\end{equation}
Analogous differential equations with respect to
other parameters can be obtained by appropriate permutation
of arguments, using the tetrahedral symmetry of the function $g$.
The solution of this differential equation is presented in
the work of Fromm and Hill in \cite{fh} in the context of the
analytic evaluation of the three-electron integral.
However, they have not used the differential equation approach, but instead
performed all integrals directly in momentum space.

\section{The master integral}
Let us assume $w_1=0$. The differential equation
in variable $t=u_1$ is
\begin{equation}
\sigma\,\frac{\partial g}{\partial t} +
\frac{1}{2}\,\frac{\partial\sigma}{\partial t}\,g + P(t,0;u_3,w_3;w_2,u_2) = 0\,,
\label{17}
\end{equation}
where $\sigma$ from Eq.(\ref{12}) is now
\begin{equation}
\sigma =
(u_2^2 - u_3^2 + w_2^2 - w_3^2)\,(u_2^2\,w_2^2 - u_3^2\,w_3^2)
- t^2\,(u_2^2 - u_3^2)\,(w_2^2 - w_3^2)\,.\label{18}
\end{equation}
This differential equation takes the form
\begin{equation}
(t^2-p^2)\,g' + t\,g = R(t),\label{19}
\end{equation}
where
\begin{eqnarray}
p^2
&=&\frac{(u_2^2-u_3^2+w_2^2-w_3^2)\,(u_2^2\,w_2^2-u_3^2\,w_3^2)}{(u_2^2-u_3^2)\,(w_2^2-w_3^2)},
\label{20}\\
R(t) &=& \frac{P(t,0;u_3,w_3;w_2,u_2)}{(u_2^2 - u_3^2)\,(w_2^2 - w_3^2)}\nonumber\\
 &=&\frac{1}{2}\,\Bigl[a^+\,R_1(t)+b^+\,R_2(t) + a^-\,R_3(t) + b^-\,R_4(t)\Bigr],\label{21}
\end{eqnarray}
and
\begin{eqnarray}
a^\pm &=& \frac{w_2}{w_2^2-w_3^2}\pm\frac{u_3}{u_2^2-u_3^2},\label{22}\\
b^\pm &=& \frac{u_2}{u_2^2-u_3^2}\pm\frac{w_3}{w_2^2-w_3^2},\label{23}\\
R_1(t) &=& \frac{1}{t+u_3-w_2}\,\ln\biggl(\frac{t+u_3+w_3}{w_2+w_3}\biggr)
-\frac{1}{t-u_3+w_2}\,\ln\biggl(\frac{t+u_2+w_2}{u_2+u_3}\biggr),\label{24}\\
R_2(t) &=& \frac{1}{t-u_2+w_3}\,\ln\biggl(\frac{t+u_3+w_3}{u_2+u_3}\biggr)
-\frac{1}{t+u_2-w_3}\,\ln\biggl(\frac{t+u_2+w_2}{w_2+w_3}\biggr),\label{25}\\
R_3(t) &=&
\frac{1}{t-u_3-w_2}\,\ln\biggl(\frac{t+u_2+w_3}{u_2+u_3+w_2+w_3}\biggr)\nonumber \\&&
-\frac{1}{t+u_3+w_2}\,\ln\biggl[
\frac{(t+u_2+w_2)\,(t+u_3+w_3)\,(u_2+u_3+w_2+w_3)}{(t+u_2+w_3)(u_2+u_3)\,(w_2+w_3)}\biggr],\label{26}\\
R_4(t) &=&
\frac{1}{t-u_2-w_3}\,\ln\biggl(\frac{t+u_3+w_2}{u_2+u_3+w_2+w_3}\biggr)\nonumber \\&&
-\frac{1}{t+u_2+w_3}\,\ln\biggl[
\frac{(t+u_2+w_2)\,(t+u_3+w_3)\,(u_2+u_3+w_2+w_3)}{(t+u_3+w_2)(u_2+u_3)\,(w_2+w_3)}\biggr].\label{27}
\end{eqnarray}
One performs the inverse Laplace transform of Eq. (\ref{19}) and obtains
the differential equation for the function $f$
\begin{equation}
r\,f''(r) +f'(r) -p^2\,r f(r)+F(r) =0\,,\label{28}
\end{equation}
where
\begin{eqnarray}
F(r) &=&
\frac{1}{2\,\pi\,i}\,\int_{-i\,\infty+\epsilon}^{i\,\infty+\epsilon}dt\,e^{t\,r}\,R(t)\label{29}\\
 &=& \frac{1}{2}\,\Bigl[a^+\,F_1^-(r)+b^+\,F_2^-(r) + a^-\,F_3^-(r) + b^-\,F_4^-(r)\Bigr],\label{30}
\end{eqnarray}
and $F^-_i$ are the inverse Laplace transform of $R_i$,
\begin{eqnarray}
F_1^\pm(r) &=& e^{r\,(u_3-w_2)}\,{\rm Ei}[-r(u_2+u_3)]\pm e^{r\,(w_2-u_3)}\,{\rm Ei}[-r(w_2+w_3)],\label{31}\\
F_2^\pm(r) &=& e^{r\,(w_3-u_2)}\,{\rm Ei}[-r(w_2+w_3)] \pm e^{r\,(u_2-w_3)}\,{\rm Ei}[-r(u_2+u_3)],\label{32}\\
F_3^\pm(r) &=& e^{-r\,(u_3+w_2)}\,\biggl\{
\ln\biggl|\frac{(u_2+w_3-u_3-w_2)(u_2+u_3)\,(w_2+w_3)}{(u_2+u_3+w_2+w_3)(u_2-u_3)(w_2-w_3)}\biggr|
-{\rm Ei}[r(u_3+w_2-u_2-w_3)]
\nonumber \\ &&
+{\rm Ei}[r(w_2-w_3)]+{\rm Ei}[r(u_3-u_2)]\biggr\}\pm e^{r\,(u_3+w_2)}\,{\rm Ei}[-r(u_2+u_3+w_2+w_3)],
\label{33}\\
F_4^\pm(r) &=& e^{-r\,(u_2+w_3)}\biggl\{
\ln\biggl|\frac{(u_2+w_3-u_3-w_2)(u_2+u_3)\,(w_2+w_3)}{(u_2+u_3+w_2+w_3)(u_2-u_3)(w_2-w_3)}\biggr|
-{\rm Ei}[r(u_2+w_3-u_3-w_2)]
\nonumber \\ &&
+{\rm Ei}[r(w_3-w_2)]+{\rm Ei}[r(u_2-u_3)]\biggr\} \pm e^{r\,(u_2+w_3)}\,{\rm Ei}[-r(u_2+u_3+w_2+w_3)].
\label{34}
\end{eqnarray}
The solution of the differential equation (\ref{28}) is
\begin{equation}
f(r) = I_0(p\,r)\int_r^\infty dr'\,F(r')\,K_0(p\,r')+K_0(p\,r)\int_0^r dr'\,F(r')\,I_0(p\,r')\,,
\label{35}
\end{equation}
where $I_0$ and $K_0$ are modified Bessel functions. This is our principal
result for the master integral $f$. In this work we do not present any numerical
examples for validation of Eq. (\ref{35}), but nevertheless verify,
that in the limit of small $r$ the function $f(r)$
\begin{equation}
f(r) = -r\, F(0) + O(r^2),
\end{equation}
where
\begin{equation}
F(0) = \frac{1}{w_2+w_3}\,\ln\biggl(\frac{u_2+u_3}{u_2+u_3+w_2+w_3}\biggr) +
       \frac{1}{u_2+u_3}\,\ln\biggl(\frac{w_2+w_3}{u_2+u_3+w_2+w_3}\biggr)
\end{equation}
coincides with the corresponding helium integral.

We will show in next sections that all the two-electron two-center integrals
can be expressed in terms of $f$, the first derivative $f'$
\begin{equation}
f'(r) = p\,\Bigl[I_1(p\,r)\int_r^\infty dr'\,F(r')\,K_0(p\,r')
-K_1(p\,r)\int_0^r dr'\,F(r')\,I_0(p\,r')\Bigr],\label{36}
\end{equation}
the exponential integral Ei, and exponential functions. In the
derivation of integrals with powers of $r_{12}$, we will need higher
order derivatives $f^{(n)}$ and they can be obtained directly from the
differential equation (\ref{28}),
\begin{equation}
f^{(n)}(r) = (n-2)\,\frac{p^2}{r}\,f^{(n-3)}(r) + p^2\,f^{(n-2)}(r)
- (n-1)\,\frac{1}{r}\,f^{(n-1)}(r) - \frac{1}{r}\,F^{(n-2)}(r).\label{37}
\end{equation}

\section{Powers of $r_{12}$}
We now pass to the calculation of $f_n(r)$, the integral with
$r_{12}^{n-1}$. For this we use two differential equations with respect to
$w_1$ and $u_1$
\begin{eqnarray}
{\rm eq}_1 &\equiv& \sigma\,\frac{\partial g}{\partial u_1} +
\frac{1}{2}\,\frac{\partial\sigma}{\partial u_1}\,g + P(u_1,w_1;u_3,w_3;w_2,u_2) = 0\,,\label{38}\\
{\rm eq}_2 &\equiv&\sigma\,\frac{\partial g}{\partial w_1} +
\frac{1}{2}\,\frac{\partial\sigma}{\partial w_1}\,g + P(w_1,u_1;w_2,u_2;w_3,u_3) = 0\,.\label{39}
\end{eqnarray}
In the first step the first equation is differentiated $n+1$ times
with respect to $w_1$ and the second equation $n$ times, at $w_1=0$. In
the second step, an inverse Laplace transform is performed of both
equations. According to Eq. (\ref{06})
\begin{equation}
f_n(r) = (-1)^n\,\frac{\partial^n}{\partial w_1^n}\biggr|_{w_1=0}\;
\frac{1}{2\,\pi\,i}\,\int_{-i\,\infty+\epsilon}^{i\,\infty+\epsilon}dt\,e^{t\,r}\,g(t)\,,\label{40}
\end{equation}
and let us introduce analogous notation
\begin{eqnarray}
U_n(r) &=&(-1)^n\,\frac{\partial^n}{\partial w_1^n}\biggr|_{w_1=0}\;
\frac{1}{2\,\pi\,i}\,\int_{-i\,\infty+\epsilon}^{i\,\infty+\epsilon}dt\,e^{t\,r}\,P(t,w_1;u_3,w_3;w_2,u_2),
\label{41}\\
W_n(r) &=& (-1)^n\,\frac{\partial^n}{\partial
w_1^n}\biggr|_{w_1=0}\;
\frac{1}{2\,\pi\,i}\,\int_{-i\,\infty+\epsilon}^{i\,\infty+\epsilon}dt\,e^{t\,r}\,P(w_1,t;w_2,u_2;w_3,u_3),
\label{42}\\
V_n(r) &=& (-1)^n\,\frac{\partial^n}{\partial
w_1^n}\biggr|_{w_1=0}\;
\frac{1}{2\,\pi\,i}\,\int_{-i\,\infty+\epsilon}^{i\,\infty+\epsilon}dt\,e^{t\,r}\,P(w_3,u_3;w_2,u_2;w_1,t).
\label{43}
\end{eqnarray}
In the third step the combination
\begin{equation}
\frac{\partial}{\partial r}({\rm eq}_1+ r\,{\rm eq}_2)+{\rm eq}_2\label{44}
\end{equation}
is formed, where all derivatives of $f_{n+1}(r)$ cancel out, and the resulting
equation is solved against $f_{n+1}(r)$
\begin{eqnarray}
f_{n+1}(r) &=& \frac{1}{(u_2^2 - u_3^2 + w_2^2 - w_3^2)\,(u_2^2\,w_2^2 - u_3^2\,w_3^2)}\biggl\{
2\,W_n(r) + r\,W'_n(r) - U'_{n+1}(r)
\nonumber \\ &&
+ n\,(u_3^2 - w_2^2)\,(u_2^2 - w_3^2)\,\bigl[r\,f'_{n-1}(r) - (n-1)\,f_{n-1}(r)\bigr]
\nonumber \\ &&
- n\,(u_2^2 + u_3^2 + w_2^2 + w_3^2)\,\bigl[r\,f^{(3)}_{n-1}(r) + 2\,f^{(2)}_{n-1}(r)\bigr]
\nonumber \\ &&
+ n\,\bigl[r\,f^{(5)}_{n-1}(r) + (n+3)\,f^{(4)}_{n-1}(r)\bigr]
\nonumber \\ &&
+ 2\,(n-2)\,(n-1)\,n\,\bigl[r\,f^{(3)}_{n-3}(r) + 2\,f^{(2)}_{n-3}(r) \bigr]\biggr\}.\label{45}
\end{eqnarray}
This recursion relation allows one to obtain integral with an
arbitrary power $n$ of $r_{12}$, knowing integrals with $n-2, n-4$
and its derivatives with respect to $r$, for example
\begin{eqnarray}
f_2(r) &=&
\frac{(u_2^2-u_3^2+w_3^2-w_2^2)\,(u_2^2\,w_3^2 - u_3^2\,w_2^2)\,f(r)}{q^2}
- \frac{a^-\,a^+\,b^-\,b^+\,q\,r\,f'(r)}{p^2}
 + \frac{f''(r)}{q}
\nonumber \\ &&
+\biggl(r+\frac{u_2}{u_2^2 - u_3^2} + \frac{w_2}{w_2^2 - w_3^2}\biggr)\,\frac{e^{-r\,(u_2 + w_2)}}{q\,r}
+ \biggl(r-\frac{u_3}{u_2^2 - u_3^2} - \frac{w_3}{w_2^2 - w_3^2}\biggr)\,\frac{e^{-r\,(u_3 + w_3)}}{q\,r}
\nonumber \\ &&
-b^-\,\frac{e^{-r\,(u_3 + w_2)}}{q\,r} - a^-\,\frac{e^{-r\,(u_2 + w_3)}}{q\,r}
-\frac{a^+\,a^-\,b^+\,b^-}{2\,p^2}\,\biggl[
\frac{(w_2 - u_3)}{a^+}\,F_1^+(r)
\nonumber \\ &&
+\frac{(u_2 - w_3)}{b^+}\,F_2^+(r)
+ \frac{(w_2 + u_3)}{a^-}\,F_3^+(r)
+ \frac{(u_2 + w_3)}{b^-}\,F_4^+(r)\biggr],\label{46}
\end{eqnarray}
where
\begin{eqnarray}
q &=& (u_2^2-u_3^2)\,(w_2^2-w_3^2).\label{47}
\end{eqnarray}

\section{Powers of $r_{1A}, r_{1B}, r_{2A}$, and $r_{2B}$}
Finally we pass to integrals with powers of $r_{1A}, r_{1B}, r_{2A}$,
and $r_{2B}$. These are obtained by differentiation of $f_n(r)$ with respect
to corresponding parameters $u_3, u_2, w_2$, and $w_3$.
Let us consider differentiation of $f$ with respect to $w_3$. We again use
differential equations to derive corresponding recursion relations, namely
\begin{eqnarray}
{\rm eq}_1 &\equiv& \sigma\,\frac{\partial g}{\partial u_1} +
\frac{1}{2}\,\frac{\partial\sigma}{\partial u_1}\,g + P(u_1,w_1;u_3,w_3;w_2,u_2) = 0\,,\label{48}\\
{\rm eq}_3 &\equiv&\sigma\,\frac{\partial g}{\partial w_3} +
\frac{1}{2}\,\frac{\partial\sigma}{\partial w_3}\,g + P(w_3,u_3;w_2,u_2;w_1,u_1) = 0\,.\label{49}
\end{eqnarray}
In the first step we differentiate both equations $n$ times with
respect to $w_1$ and set $w_1=0$. In the second step we perform an
inverse Laplace transform. In the third step we form the expression
\begin{equation}
\frac{\partial}{\partial r}\biggl(\frac{\partial{\rm eq}_1}{\partial w_3}  +
r\,{\rm eq}_3\biggr) + {\rm eq}_3\,, \label{50}
\end{equation}
which cancels out derivatives of $f_n$ with respect to $t$, and solve the
corresponding equation against $f_n(0,0,0,1;r)$
\begin{eqnarray}
f_n(0,0,0,1;r) &\equiv& -\frac{\partial f_n }{\partial w_3}\nonumber
\\ &=& \frac{1}{(u_2^2 - u_3^2 + w_2^2 - w_3^2)\,(u_2^2\,w_2^2 -
u_3^2\,w_3^2)} \biggl[2\,V_n(r) +r\,V'_n(r) + \frac{\partial
U_n(r)}{\partial w_3} \nonumber \\ && + w_3\,(u_2^2\,u_3^2 - u_3^4 +
u_2^2\,w_2^2 + u_3^2\,w_2^2 - 2\,u_3^2\,w_3^2)\,r\,f'_n(r) \nonumber
\\ && - w_3\,(u_2^2-u_3^2)\,\bigl[2\,f^{(2)}_n(r) +
r\,f^{(3)}_n(r)\bigr] \nonumber \\ && + (n-1)\,n\,\biggl[(u_3^2 -
w_2^2)\,(u_2^2 - w_3^2)\,\frac{\partial f_{n-2}(r)}{\partial w_3} -
\frac{\partial f^{(4)}_{n-2}(r)}{\partial w_3} \nonumber \\ && +
r\,(u_3^2 - w_2^2)\,w_3\,f'_{n-2}(r) + 2\,w_3\,f^{(2)}_{n-2}(r) +
r\,w_3\,f^{(3)}_{n-2}(r) \biggr]\biggr].\label{51}
\end{eqnarray}
 In the particular case of $n=0$ it takes the form
\begin{eqnarray}
f(0,0,0,1;r) &=& -\frac{\partial f}{\partial w_3} \nonumber \\ &=&
- \frac{w_3}{w_2^2-w_3^2}\,f(r) - \frac{a^-\,a^+\,w_3\,(u_2^2-u_3^2)}{p^2}\,r\,f'(r)\nonumber \\&&
+\frac{1}{2\,p^2\,(w_2^2-w_3^2)}\,\biggl\{
  a^-\,(u_3 - w_2)\,w_3\,F_1^+
- a^+\,(u_3 + w_2)\,w_3\,F_3^+ \nonumber \\ && -
\frac{1}{q}\,[u_2\,w_2^2\,(u_2^2 - u_3^2 + w_2^2 - w_3^2) -
w_3\,(u_2^2\,w_2^2 - u_3^2\,w_3^2)]\,F_2^+ \nonumber \\ && +
\frac{1}{q}\,[u_2\,w_2^2\,(u_2^2 - u_3^2 + w_2^2 - w_3^2) +
w_3\,(u_2^2\,w_2^2 - u_3^2\,w_3^2)]\,F_4^+\biggr\}.\label{52}
\end{eqnarray}
The other single powers of the electron distances can be obtained from the above
by appropriate exchange of $u_2,u_3,w_2$, and $w_3$.
The general recursion can be obtained by further
differentiation of Eq. (\ref{51}) (after multiplying by the common denominator)
with respect to $w_2, u_2, w_3, u_3$,
or by recursive application of this single differentiation formulae.

\section{ Special case: atomic orbitals}
The explicit form of $f_n(i,j,k,l;r)$ in the general case becomes 
very lengthy for increasing values of $i,j,k$, and $l$. Therefore
it is worth while to consider special cases which may find practical realization
in quantum chemistry codes. 
When in Eq. (\ref{06}) atomic orbitals are used, then
two nonlinear parameters in the direct (no exchange) integral are $u_2 = w_2=0$.
We thus assume here vanishing of $u_2$ and $w_2$, but allow for an arbitrary polynomial
in electron-nucleus and electron-electron distances.
If we introduce the notation $u_3 = u$, and $w_3 = w$, then $p = \sqrt{u^2+w^2}$, and
\begin{eqnarray}
f(r) &=& \int \frac{d^3 r_1}{4\,\pi}\,\int \frac{d^3 r_2}{4\,\pi}\,
\frac{e^{-u\,r_{1A}}}{r_{1A}}\,\frac{e^{-w\,r_{2B}}}{r_{2B}}\,
\frac{1}{r_{1B}}\,\frac{1}{r_{2A}}\,\frac{r}{r_{12}}\label{53}\\
&=& I_0(p\,r)\int_r^\infty dr'\,F(r')\,K_0(p\,r')+K_0(p\,r)\int_0^rdr'\,F(r')\,I_0(p\,r')\,,\label{54}
\end{eqnarray}
and $F$ becomes
\begin{eqnarray}
F(r) &=&
-\frac{e^{u\,r}}{2\,u}\Bigl[{\rm Ei}\bigl(-r\,(u+w)\bigr) + {\rm Ei}(-r\,u)\Bigr]
-\frac{e^{w\,r}}{2\,w}\Bigl[{\rm Ei}\bigl(-r\,(u+w)\bigr) + {\rm Ei}(-r\,w)\Bigr]
\nonumber \\ &&
+\frac{e^{-u\,r}}{2\,u}\biggl[\ln\biggl|\frac{w-u}{w+u}\biggr|
-{\rm Ei}\bigl(r(u-w)\bigr)+{\rm Ei}(r\,u)+2\,{\rm Ei}(-r\,w)\biggr]
\nonumber \\ &&
+\frac{e^{-w\,r}}{2\,w}\biggl[\ln\biggl|\frac{w-u}{w+u}\biggr|
-{\rm Ei}\bigl(r(w-u)\bigr)+{\rm Ei}(r\,w)+2\,{\rm Ei}(-r\,u)\biggr]\,.\label{55}
\end{eqnarray}
All recursion formulae for higher powers of electron distances
can be obtained directly from the general case
considered in the previous section by setting $u_2=w_2=0$, and
they take here a much simpler form.

\section{ Special case: exchanged atomic orbitals}
For the exchange integral with atomic orbitals, the relation
$u_3 = w_2 \equiv u, u_2 = w_3 \equiv w$ holds, then $p=0$ and using the
small $x$ limit of Bessel functions,
\begin{eqnarray}
{\rm I}_0(x) &=& 1+ O(x)\,,\label{56}\\
{\rm K}_0(x) &=& -\biggl(\gamma+\ln\frac{x}{2}\biggr) + O(x)\,.\label{57}
\end{eqnarray}
the master integral becomes
\begin{eqnarray}
f(r) &=& \int \frac{d^3 r_1}{4\,\pi}\,\int \frac{d^3 r_2}{4\,\pi}\,
\frac{e^{-u\,(r_{1A}+r_{2A})}}{r_{1A}\,r_{2A}}\,
\frac{e^{-w\,(r_{1B}+r_{2B})}}{r_{1B}\,r_{2B}}\,
\frac{r}{r_{12}}\label{58}\\
&=& \int_r^\infty dr'\,F(r')\,\ln\frac{r}{r'}
=  \int_r^\infty dr'\,F^{(-1)}(r')\,\frac{1}{r'}\,,\label{59}
\end{eqnarray}
where
\begin{eqnarray}
F^{(-1)}(r) &=& \frac{1}{2\,(u^2 - w^2)}\,
\biggl\{(e^{2\,r\,w}-e^{2\,r\,u})\,{\rm Ei}(-2\,r\,(u + w))\label{60}
\\&&
- e^{-2\,r\,w}\,\biggl[{\rm Ei}(2\,r\,(w - u))
- 2\,{\rm Ei}(r\,(w - u)) + \ln\biggl|\frac{w-u}{w+u}\biggr|\biggr]
\nonumber \\ &&
+ e^{-2\,r\,u}\,\biggl[{\rm Ei}(2\,r\,(u - w))
- 2\,{\rm Ei}(r\,(u - w)) + \ln\biggl|\frac{w-u}{w+u}\biggr|\biggr]\biggr\},
\nonumber
\end{eqnarray}
and the superscript $^{(-1)}$ denotes not the differentiation, but the integration
over $r$ with the boundary condition $F^{(-1)}(\infty)=0$, so
$dF^{(-1)}(r)/dr = F(r)$.

The recurrence relations for powers of $r_{12}$ is obtained
from the differential equation in $w_1$, Eq. (\ref{39}).
One divides it by $t^2$, differentiates over $w_1$ $n$-times,
performs the inverse Laplace transform, and obtains
\begin{eqnarray}
f_{n+1}(r) &=&\frac{1}{(u^2 - w^2)^2}\,\bigl[
2\,n^2\,(u^2 + w^2)\,f_{n-1}(r)
- n^2\,f^{(2)}_{n-1}(r)
-(n-2)\,(n-1)^2\,n\,f_{n-3}(r)
\nonumber \\ &&
 + W_n^{(-2)}(r) \bigr]\,,\label{61}
\end{eqnarray}
where
\begin{eqnarray}
W_n^{(-2)}(r) &=& (-1)^n\,\frac{\partial^n}{\partial w_1^n}\biggr|_{w_1=0}\, W^{(-2)}(r)\,,\label{62} \\
W^{(-2)}(r) &=&
\frac{e^{-2\,r\,u}}{r}\,\biggl(1 - \frac{w_1}{w_1 - u + w} - \frac{w_1}{w_1 + u + w}\biggr)
\nonumber \\ &&
+ \frac{e^{-2\,r\,w}}{r}\,\biggl(1 - \frac{w_1}{w_1 + u - w} - \frac{w_1}{w_1 + u + w}\biggr)
\nonumber \\ &&
+ \frac{e^{-r\,(w_1+u+w)}}{r}\,\biggl(-2 + \frac{w_1}{w_1 -u + w}
+ \frac{w_1}{w_1 + u - w} + \frac{2\,w_1}{w_1 + u + w}\biggr)
\nonumber \\ &&
- 2\,w_1\,{\rm Ei}(-r\,(u + w + w_1))\,.\label{63}
\end{eqnarray}
Similarly differentiation of the master integral
with respect to nonlinear parameter, for example $w_3$
is obtained from the differential equation in this nonlinear parameter
\begin{eqnarray}
f(0,0,0,1;r) &=& \frac{w}{u^2-w^2}\,f(r)
+ \frac{1}{4\,w\,(u^2-w^2)}\,\biggl\{ 2\,{\rm Ei}(-r\,(u + w)) - e^{2\,r\,w}\,{\rm Ei}(-2\,r\,(u + w))
\nonumber \\ &&
+ e^{-2\,r\,w}\,\biggl[{\rm Ei}(2\,r\,(w - u)) - 2\,{\rm Ei}(r\,(w - u))
+ \ln\biggl|\frac{w-u}{w+u}\biggr|\biggr]\biggr\}
\label{64}
\end{eqnarray}
The appearance of $u-w$ in the denominator affects numerical stability of
these recursions when $u\approx w$. This problem can be probably solved by
employing sufficiently lengthy Taylor expansions around $u=w$,
and in this special case $f$ is known analytically, as will be discussed in the next section.

\section{ Special case: symmetric}
This is the case when nonlinear parameters are the same for nuclei
$A$ and $B$, namely $w_2=w_3 = w, u_2 = u_3 = u$, and then
$p=\infty$. It is the James-Coolidge basis for H$_2$ molecule, and
was recently used by Sims and Hagstrom \cite{SH06} for the very
accurate calculation of BO potential for small nuclear distances.
Here we show that all integrals can be expressed in terms of Ei and
exponential functions. It is
convenient in this case to consider a slightly different form of the
integral, namely
\begin{eqnarray}
f(n_1,n_2,n_3,n_4,n_5;r) &=& \int \frac{d^3 r_1}{4\,\pi}\,\int \frac{d^3 r_2}{4\,\pi}\,
\frac{e^{-u\,r_{1A}}}{r_{1A}}\,
\frac{e^{-u\,r_{1B}}}{r_{1B}}\,
\frac{e^{-w\,r_{2A}}}{r_{2A}}\,
\frac{e^{-w\,r_{2B}}}{r_{2B}}\,\frac{r}{r_{12}^{1-n_1}}\nonumber \\ &&
(r_{1A}-r_{1B})^{n_2}\,(r_{2A}-r_{2B})^{n_3}\,(r_{1A}+r_{1B})^{n_4}\,(r_{2A}+r_{2B})^{n_5}.\label{65}
\end{eqnarray}
$f$ for all values of parameters $n_i$ can be obtained from one differential equation
in variable  $u_1\equiv t$
\begin{equation}
\sigma\,\frac{\partial g}{\partial t} + \frac{1}{2}\frac{\partial
  \sigma}{\partial t}\,g + P(t,w_1;u_3,w_3;w_2,u_2)=0\label{66}
\end{equation}
by the inverse Laplace transform in $t$ and differentiation with
respect to $w_1, (w_2-w_3)/2, (u_2-u_3)/2, (w_2+w_3)/2,$ and
$(u_2+u_3)/2$, at $w_1=w_2-w_3=u_2-u_3=0$. This differential
equation becomes then an algebraic equation, which relates values of
$f$ for different arguments, and can easily be solved. For example,
the master integral is
\begin{eqnarray}
f(0,0,0,0,0;r) &=&\frac{1}{4\,u\,w}\biggl[ e^{r\,(u+w)}\,{\rm
Ei}\bigl(-2\,r\,(u+w)\bigr)
+e^{-r(u+w)}\,\biggl(\gamma+\ln\frac{2\,r\,u\,w}{u+w}\biggr)
\nonumber \\ && -e^{r(u-w)}\,{\rm Ei}(-2\,r\,u) - e^{r(w-u)}\,{\rm
Ei}(-2\,r\,w)\biggr].\label{67}
\end{eqnarray}
Other examples include
\begin{eqnarray}
f(2,0,0,0,0;r)  &=&
\frac{(u^2+w^2)}{6\,u^2\,w^2}\,f(0,0,0,0,0;r) + \frac{e^{-r\,(u + w)}\,r^2}{12\,u\,w}\nonumber \\ &&
+  \frac{r}{24\,u^2\,w^2}\,\biggl[(u + w)\,e^{-r\,(u + w)}
- (u - w)\,e^{r\,(u - w)}\,{\rm Ei}(-2\,r\,u)  \nonumber \\ &&
- (w - u)\,e^{r\,(w - u)}\,{\rm Ei}(-2\,r\,w)
- (u + w)\,e^{r\,(u + w)}\,{\rm Ei}(-2\,r\,(u + w))  \nonumber \\ &&
+ (u + w)\,e^{-r\,(u + w)}\,\biggl(\gamma + \ln\frac{2\,r\,u\,w}{u + w}\biggr)\biggr]\,,\label{68}
 \\
f(0,2,0,0,0;r)  &=& \frac{r^2}{3}\,f(0,0,0,0,0;r)\,.\label{69}
\end{eqnarray}
Since all other integrals can also be expressed in terms of the Ei and exponential
functions, matrix elements of the nonrelativistic Hamiltonian
can all be obtained analytically. This should allow one to obtain 
highly accurate nonrelativistic wave functions, and thus precisely calculate
various relativistic effects to rovibrational energies, shielding 
and spin-rotational constants in the H$_2$ molecule.

\section{Summary}
We have presented an approach to evaluate two-center two-electron integrals
with exponential functions and with the arbitrary polynomial in electron-nucleus and
electron-electron distances. All integrals are expressed in terms
of the master integral $f$, the derivative $f'$, Ei and exponential functions. 
The master integral satisfies the second order differential equation
(\ref{28}) in variable $r=r_{AB}$, and can be accurately solved.
This approach certainly finds an application in the H$_2$ molecule,
for example the present theoretical predictions for the dissociation energy
\cite{wol, h2} are limited by unknown higher order $m\,\alpha^6$ corrections and 
the finite nuclear mass effects in the leading relativistic corrections $m\,\alpha^4$.
Both of these corrections are difficult (if not impossible)
to calculate using Gaussian functions.  Apart from H$_2$,
this approach may find applications in arbitrary few electron diatomic
molecules. The special cases of integrals with direct and exchange atomic
functions were considered for this purpose. We do not know however,
how well relativistic effects can be calculated in this aproach. 
This would require numerical experiments. But the message is that the integrals
with exponential functions can now be precisely calculated.

\section*{Acknowledgments}
This work was supported by NIST
through Precision Measurement Grant PMG 60NANB7D6153.

\end{document}